\documentclass[10pt]{iopart}

\usepackage{iopams}
\usepackage{amssymb}
\usepackage{graphicx}
\usepackage{txfonts}
\usepackage{url}
\usepackage{cite}
 \usepackage[colorlinks, citecolor = blue, urlcolor = blue]{hyperref}
 \usepackage{breakurl}

\newcommand{\Excursion} {{\mathbb E}}
\newcommand{\Sspace}    {{\mathbb S}}
\newcommand{\Nside}     {{N_{\hbox{\scriptsize side}}}}
\newcommand{\lmax}      {{l_{\hbox{\scriptsize max}}}}

\begin{document}

\title[Betti Functionals in Cosmic Topology]
      {Betti Functionals as a Probe for Cosmic Topology}

\author{Ralf Aurich and Frank Steiner}

\address{Ulm University, Institute of Theoretical Physics, Albert-Einstein-Allee 11, D--89069 Ulm, Germany}
\vspace{10pt}
\begin{indented}
\item[]March 2024
\end{indented}

\begin{abstract}
  The question of the global topology of the Universe (cosmic topology) is
  still open.
  In the $\Lambda$CDM concordance model it is assumed
  that the space of the Universe possesses the trivial topology
  of $\mathbb{R}^3$ and thus that the Universe has an {\it infinite} volume.
  As an alternative, we study in this paper one of the simplest non-trivial
  topologies given by a cubic 3-torus describing
  a universe with a {\it finite} volume.
  To probe cosmic topology, we analyze certain
  structure properties in the cosmic microwave background (CMB) using
  {\it Betti Functionals} and the {\it Euler Characteristic} evaluated
  on excursions sets,
  which possess a simple geometrical interpretation.
  Since the CMB temperature fluctuations $\delta T$ are observed on the sphere
  $\Sspace^2$ surrounding the observer,
  there are only three Betti functionals $\beta_k(\nu)$, $k=1,2,3$.
  Here $\nu=\delta T/\sigma_0$ denotes the temperature threshold normalized
  by the standard deviation $\sigma_0$ of $\delta T$.
  Analytic approximations of the Gaussian expectations for the
  Betti functionals and an exact formula for the Euler characteristic
  are given.
  It is shown that the amplitudes of $\beta_0(\nu)$ and $\beta_1(\nu)$
  decrease with increasing volume $V=L^3$ of the cubic 3-torus universe.
  Since the computation of the $\beta_k$'s from observational sky maps
  is hindered due to the presence of masks,
  we suggest a method yielding lower and upper bounds for them
  and apply it to four Planck 2018 sky maps.
  It is found that the $\beta_k$'s of the Planck maps lie between those
  of the torus universes with side-lengths $L=2.0$ and $L=3.0$
  in units of the Hubble length
  and above the infinite $\Lambda$CDM case.
  These results give a further hint
  that the Universe has a non-trivial topology.
 \end{abstract}

%
\vspace{2pc}
\ioptwocol
%

\section{Introduction}

A wide range of cosmological data is well described within the framework
of the $\Lambda$ Cold Dark Matter ($\Lambda$CDM) model,
which is now established as the standard model of cosmology.
Due to the increasing level of precision of the available data, however,
several discrepancies have arisen as the Hubble tension and
the $S_8$ tension in the recent years
\cite{Anchordoqui_et_al_2021, DiValentino_2021, Abdalla_et_al_2022, %
 Vagnozzi_2023, Akarsu_et_al_2024}.
In addition, there are several strange features in the
cosmic microwave background (CMB) as summarized in
\cite{Schwarz_et_al_2016},
which also point to a modification of the standard model.
A prominent example is the suppression of the quadrupole moment in the CMB
angular power spectrum,
which is also especially revealed by the 2-point angular correlation function
$C(\vartheta)$ showing almost no correlations above angles of $60^\circ$.
Models without relying on modified physics, which can address this feature,
are provided by cosmic topology \cite{Ellis_1971},
see also references in \cite{Aurich_Janzer_Lustig_Steiner_2007}.
A possible non-trivial topology for the Universe can suppress the
large scale anisotropy due to a infared cut-off in the wave number spectrum.

A non-trivial topology can be detected by searching
for topologically matched circles in CMB maps,
the so-called circles-in-the-sky (CITS) test
\cite{Cornish_Spergel_Starkman_1998b}.
However, the searches were in vain up to now
\cite{Aurich_Lustig_2013,Planck_Topo_2013}.
A possibility for this negative result might be that the CITS signature
is not so clearly pronounced in the CMB sky maps as predicted by
the $\Lambda$CDM model on which the likelihoods are based.
A future resolution of the Hubble and $S_8$ tensions might, for example,
lead to a larger integrated Sachs-Wolfe contribution which would
additionally blur the CITS signature,
so that the Universe might possess a non-trivial topology despite
the negative result.
Another possibility for a non-trivial topology is
that the topology produces CITS
which are far from antipodal as in the case of the Hantzsche-Wendt topology
\cite{Aurich_Lustig_2014} which impede the CITS search.
An exhaustive analysis of allowed non-trivial topological models is
given in \cite{Akrami_et_al_2022,Petersen_et_al_2023}.

In this paper we analyze certain structure properties in the CMB maps of
a compact non-trivial topological space and compare them
with the standard $\Lambda$CDM model which presuppose an infinite volume.
This requires the simulation of CMB sky maps for these topological models
as outlined in
\cite{Riazuelo_et_al_2004,Aurich_Lustig_2010b,Eskilt_et_al_2023}.
There it is discussed that the CMB simulations require the determination of the
eigenvalue spectrum and the eigenmodes of the Laplace-Beltrami operator on the
topological space.
The obvious restriction would be to test only the topological spaces
that are not excluded by the CITS test,
but their CMB sky maps are more difficult to simulate than the simple
cubic 3-torus model whose CITS signature could not be found.
The reason is that the highest degeneracy in the eigenvalue spectrum
belongs to the cubic 3-torus topology
such that the transfer function has to be computed for a significantly
smaller number of eigenvalues
which in turn speeds up the CMB simulations correspondingly.
For that reason we consider here the cubic 3-torus space
\cite{Aurich_Janzer_Lustig_Steiner_2007}
which should provide an idea of general properties.

In this work, we apply the {\it Betti Functionals} to the CMB temperature
fluctuation field $\delta T(\hat n)$ which is defined on the sphere $\Sspace^2$
which surrounds the observer.
($\hat n$ denotes the direction in which the temperature fluctuation
$\delta T(\hat n)$ is observed.)
For the analysis of the CMB, the temperature fluctuation field
$\delta T(\hat n)$ is normalized
\begin{equation}
\label{deltaT_normalization}
f(\hat n) \; := \;  \frac{\delta T(\hat n) \, - \, \mu}{\sigma_0}
\hspace{10pt} ,
\end{equation}
where $\mu$ and $\sigma_0$ are the mean and the standard deviation of the
field $\delta T(\hat n)$.
Therefore, the normalized field $f(\hat n)$ has zero mean and unit variance.
The topological descriptors are then computed as a functional of the
\emph{excursion set}
\begin{equation}\label{excursion_set}
\Excursion(\nu) \; = \; \{\, \hat n \in \Sspace^2  \mid  f(\hat n) \geq \nu \}
\hspace{10pt} ,
\end{equation}
where $\nu$ denotes the threshold.

For our special case $\Sspace^2$, there are only three Betti Functionals (BF)
$\beta_k(\nu)$.
These are the number of components of the excursion set by the $0$-th
BF $\beta_0$,
\begin{equation}
\label{Def_Betti0}
\beta_0(\nu) \; := \;
\# \hbox{ components of } \Excursion(\nu)
\hspace{10pt} ,
\end{equation}
the number of independent 1-dimensional punctures on $\Sspace^2$,
denoted by the $1$-st BF $\beta_1$,
\begin{equation}
\label{Def_Betti1}
\beta_1(\nu) \; := \;
\# \hbox{ topological holes of } \Excursion(\nu)
\hspace{10pt} ,
\end{equation}
and finally, the $2$-nd BF $\beta_2$
which counts the number of the internal voids of $\Excursion(\nu)$.
In our special case,
this is one if the excursion set is identical to $\Sspace^2$
being the case for $\nu < \min(f(\hat n))$ and zero otherwise, i.\,e.\
\begin{equation}
\label{Def_Betti2}
\beta_2(\nu) \; := \;
\cases{
1 & for $\nu < \min(f(\hat n))$ \\
0 & otherwise }
\hspace{10pt} .
\end{equation}

The Betti numbers originate from the analysis of topological spaces,
where they are used to distinguish topological spaces based on
the connectivity of $n$-dimensional simplicial complexes.
The $n$th Betti number is defined by the rank of the $n$th homology group
\cite{Adler_2010,Edelsbrunner_Harer_2022,Munkres_2018,Pranav_2022}.
An analysis based on relative homology with respect to cosmic topology
will be given in the companion paper
\cite{Pranav_Aurich_Buchert_France_Steiner_2024}.

Concerning the word ``topological'' in eq.\,(\ref{Def_Betti1})
there is a remark in order.
At first, consider a pixelized version of the sphere $\Sspace^2$ and identify it
with the excursion set for the case $\nu < \min(f(\hat n))$.
Then increase the threshold $\nu$ such that a single pixel is removed
from the sphere which destroys the internal void, 
which in turn is counted by $\beta_2$,
but this does not create a 1-dimensional hole.
The 1-dimensional hole is created on the sphere after removal of another pixel
not sharing a boundary with the first removed pixel.
In this case we now have a ring, that has an actual hole,
which is called a topological hole.
Thus, the number of topological holes is the number of common sense holes
minus one.

In this paper we use for the $\Lambda$CDM concordance model
the cosmological parameters as given by the Planck Collaboration
in \cite{Planck_2015_XIII}
in their table 4 in the column 'TT+lowP+lensing'.
The main parameters are $\Omega_{\hbox{\scriptsize b}}h^2 =0.02226$,
$\Omega_{\hbox{\scriptsize c}}h^2 = 0.1186$, and $h=0.678$.
Furthermore, all sky maps are computed with the Healpix resolution
$N_{\hbox{\scriptsize side}}=128$ and,
if not explicitly stated otherwise, a Gaussian smearing of
$120\hbox{ arcmin}$ is used.
For the cubic 3-torus model, 1000 CMB sky maps are computed for each of the
side-lengths $L = 0.5$, $1.0$, $1.5$, $2.0$, and $3.0$,
where $L$ is given in units of the Hubble length $L_H=c/H_0$.
This allows the computation of ensemble averages for the $\beta$'s
in order to compare them with the infinite $\Lambda$CDM standard model
and the Planck CMB maps.

\section{Analytic Approximation of Betti Functionals and an exact Gaussian
         expectation for the Euler characteristic and the genus}
\label{sec:analytic_approximation}

After many years of work on the Betti functionals (BF)
$\beta_k(\nu)$ $(k=0,1,2)$,
there still exist no exact predictions for the ensemble expectations assuming
that the CMB anisotropy is a homogeneous, isotropic Gaussian random field on the
${\mathbb S}^2$-sphere.
There arises therefore the question whether one can find reliable analytic
approximations for the $\beta_k(\nu)$ as a function of $\nu$.
In the following we propose such approximations which agree well with the
average values obtained from simulations for a cubic 3-torus with
side-length $L=2.0$.

\begin{figure}
\includegraphics[width=0.5\textwidth]{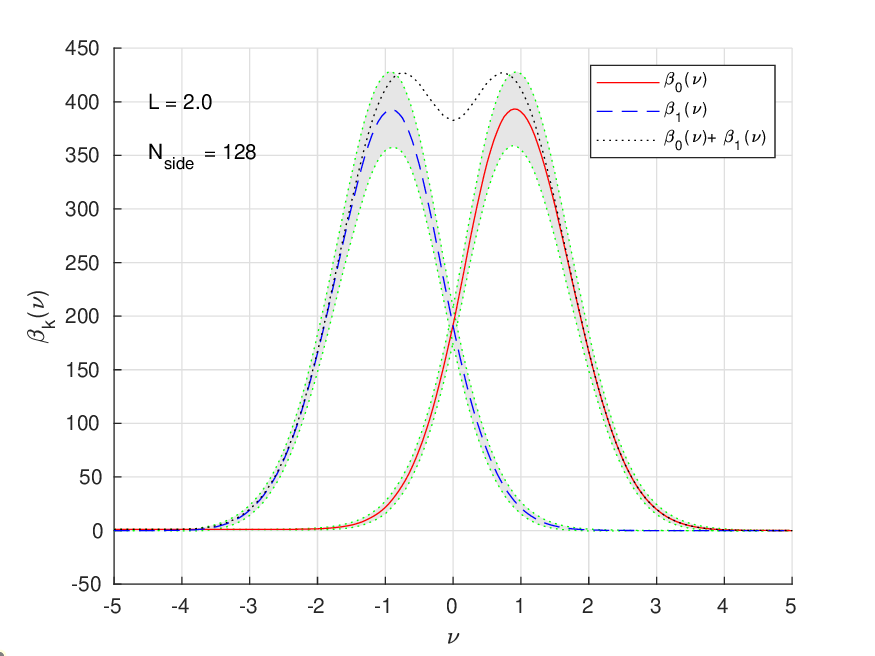}
\caption{\label{fig:beta_even}
  The Betti Functionals $\beta_0(\nu)$ and $\beta_1(\nu)$ are shown together
  with their $1\sigma$ error bands obtained from 1000 CMB simulations based on
  the torus topology.
  In addition, the mean value of $\beta_0(\nu)+\beta_1(\nu)$ is plotted
  as a dotted curve.
  The resolution parameters are set to $N_{\hbox{\scriptsize side}}=128$,
  $\lmax =256$ and $\hbox{FWHM}=120\hbox{ arcmin}$.
  These parameters are used in all figures if not otherwise noted.
}
\end{figure}

A crucial r\^ole is played by the asymptotic behaviour of the BFs in the limits
$\nu \to \pm\infty$.
From the definition of the BFs it is obvious that it holds
\begin{equation}
\label{beta_plus_infty}
\beta_k(\nu) \; \to \; 0
\hspace{10pt} \hbox{for} \hspace{10pt}
\nu \to \infty
\hspace{10pt} , \hspace{10pt}
k =0,1,2 \hspace{10pt} .
\end{equation}
The situation is, however, different in the limit $\nu \to -\infty$
where the excursion set covers the whole unit sphere ${\mathbb S}^2$.
In this case one obtains from the {\it Betti numbers} $p_k$
of ${\mathbb S}^2$ the asymptotic behaviour for $\nu \to -\infty$
\begin{eqnarray}
\beta_0(\nu) & \to & p_0 \; = \; 1 \nonumber \\
\beta_1(\nu) & \to & p_1 \; = \; 0 \label{beta_minus_infty} \\
\beta_2(\nu) & \to & p_2 \; = \; 1 \nonumber
\hspace{10pt} .
\end{eqnarray}
This is in agreement with the general relation for the Euler characteristic (EC)
$\chi(\nu)$,
\begin{equation}
\label{Euler_beta}
\chi(\nu) \; = \; \sum_{k=0}^2 (-1)^k\; \beta_k(\nu)
\hspace{10pt} ,
\end{equation}
yielding for the full sphere the correct value $\chi(-\infty)=2$.

\begin{figure}
\includegraphics[width=0.5\textwidth]{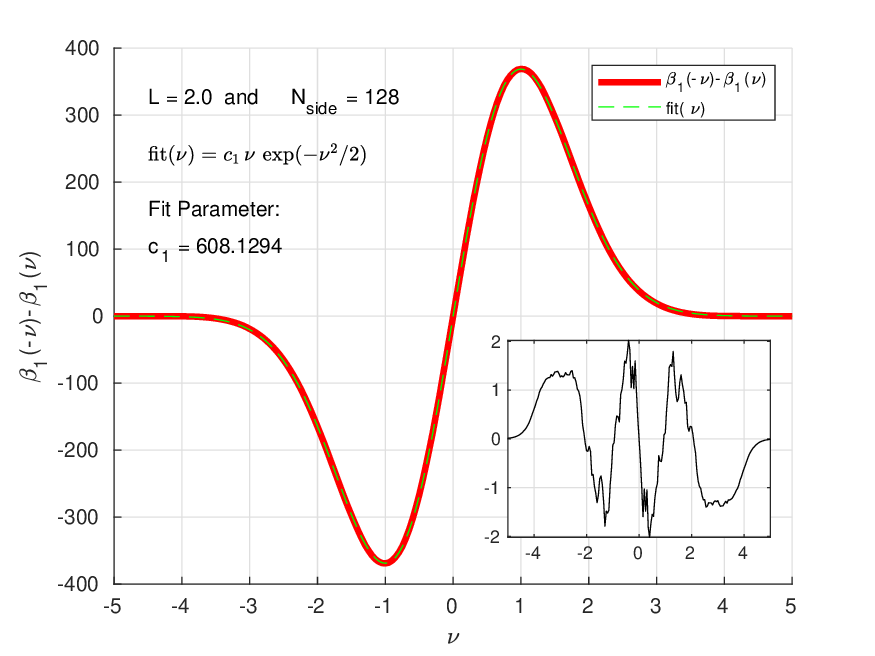}
\caption{\label{fig:beta_odd}
  The fit (\ref{coeff_c1}) is shown as a dashed curve together with the
  mean value of $\beta_1(-\nu)-\beta_1(\nu)$ obtained from 1000 CMB
  simulations based on the torus topology.
  The quality of the fit is so good that the mean value curve
  is plotted with a larger line width in order to reveal the dashed curve
  belonging to the one-parameter fit.
  The inset shows the difference between the data and the fit.
}
\end{figure}

From (\ref{beta_plus_infty}) and (\ref{beta_minus_infty}) one infers
that the apparent symmetry (see Figs.\ \ref{fig:beta_even},
\ref{fig:beta_fwhm_sequence} and \ref{fig:beta_L_sequence}),
$\beta_0(\nu) = \beta_1(-\nu)$, with respect to the parity transformation
$\nu\to-\nu$ ($\delta T \to -\delta T$ respectively),
does not hold true but rather is violated.
The degree of the breaking of parity can be described by a function
$\alpha_0(\nu)$,
\begin{equation}
\label{parity_violation}
\beta_0(\nu) \; = \; \beta_1(-\nu) \; + \; \alpha_0(\nu)
\end{equation}
which according to the Eqs.\,(\ref{beta_plus_infty}) and
(\ref{beta_minus_infty}) has to satisfy the asymptotic relations
\begin{eqnarray}
\alpha_0(\nu) & \to & 0
\hspace{10pt} \hbox{for} \hspace{10pt}
\nu \to \infty
\nonumber \\
\alpha_0(\nu) & \to & 1
\hspace{10pt} \hbox{for} \hspace{10pt}
\nu \to -\infty
\label{alpha_asymp}
\hspace{10pt} .
\end{eqnarray}
The relation (\ref{Euler_beta}) for the EC is then
\begin{equation}
\label{Euler_alpha}
\chi(\nu) \; = \;
\big( \beta_1(-\nu)-\beta_1(\nu) \big) \; + \; \alpha_0(\nu) + \beta_2(\nu)
\hspace{10pt} .
\end{equation}
Under very general assumptions, one can expand the expression in the brackets of
(\ref{Euler_alpha}) into a convergent infinite series in terms of the
odd Hermite functions
$\varphi_{2n+1}(\nu) \sim \exp(-\nu^2/2) \hbox{ He}_{2n+1}(\nu)$,
where $\hbox{He}_{2n+1}(\nu)$ denote the odd ``probabilist's'' Hermite
polynomials (see Appendix C in \cite{Buchert_France_Steiner_2017}
for a mathematical exposition of the general Hermite expansions).
As a first approximation let us consider the first term $(n=0)$
in this expansion
which leads with $\hbox{He}_{1}(\nu)=\nu$ to the approximation
\begin{equation}
\label{coeff_c1}
\beta_1(-\nu)-\beta_1(\nu) \; = \; c_1 \, \nu \, e^{-\nu^2/2}
\end{equation}
with a positive coefficient given below.
In Fig.\,\ref{fig:beta_odd} we show that (\ref{coeff_c1}) gives an
excellent fit to the data.
Eq.\,(\ref{coeff_c1}) describes a maximum at $\nu=1$ and
a minimum at $\nu=-1$ with amplitude $\pm c_1/\sqrt{e}$ in nice agreement
with the peaks of the data very close to
$\pm 1$ (see Fig.\,\ref{fig:beta_odd}).
Inserting relation (\ref{coeff_c1}) into (\ref{Euler_alpha})
one obtains the following approximation to the EC
\begin{equation}
\label{chi_approx}
\chi(\nu) \; = \;
c_1 \, \nu \, e^{-\nu^2/2} \, + \, \big(\alpha_0(\nu) + \beta_2(\nu)\big)
\hspace{10pt} .
\end{equation}
In order to determine $c_1$ and $\alpha_0(\nu)$ we use the standard definition
of the EC using the Gauss-Bonnet theorem on the excursion set
\begin{equation}
\label{Gauss-Bonnet}
\int_{{\mathbb{E}}(\nu)} K \, da \, + \, \int_{\partial{\mathbb{E}}(\nu)} \kappa(s) \, ds
\; = \; 2\pi \, \chi(\nu)
\hspace{10pt} ,
\end{equation}
where $K=1$ denotes the Gaussian curvature of ${\mathbb S}^2$,
$da$ the surface element on ${\mathbb S}^2$, $ds$ the line element along
$\partial{\mathbb{E}}(\nu)$, and $\kappa(s)$ the geodesic curvature of
$\partial{\mathbb{E}}(\nu)$.
The integrals in (\ref{Gauss-Bonnet}) are proportional to the
{\it Minkowski functionals} (MFs)
$\varv_0(\nu)$ and $ \varv_2(\nu)$ \cite{Buchert_France_Steiner_2017},
respectively, which gives
\begin{equation}
\label{chi_MF}
\chi(\nu) \; = \; 2 \, \varv_0(\nu) \, + \, 4\pi\,  \varv_2(\nu)
\hspace{10pt} .
\end{equation}
The MFs have the nice property that their exact Gaussian predictions are
explicitly known (see \cite{Buchert_France_Steiner_2017} and
references therein),
\begin{eqnarray}
\nonumber
\varv_0(\nu) & = & \frac 12 \hbox{ erfc}\left(\frac\nu{\sqrt{2}}\right) \\
\label{Minkowski_Gauss}
\varv_2(\nu) & = & \frac{\rho^2}{2(2\pi)^{3/2}} \, \nu \, \exp(-\nu^2/2)
\hspace{10pt} .
\end{eqnarray}
Here the parameter $\rho := \sigma_1/\sigma_0$ has been studied
in \cite{Aurich_Buchert_France_Steiner_2021},
where it has been shown that $\rho$ does hierarchically detect the change in
size of the cubic 3-torus, if the volume of the Universe is smaller than
$\simeq 2.5 \times 10^3 \hbox{Gpc}^3$.
($\sigma_1$ is the standard deviation of the gradient of the CMB field
$\delta T(\hat n)$, i.\,e.\ $\rho$ is the normalized standard deviation
of the CMB gradient field.)

\begin{figure}
\includegraphics[width=0.5\textwidth]{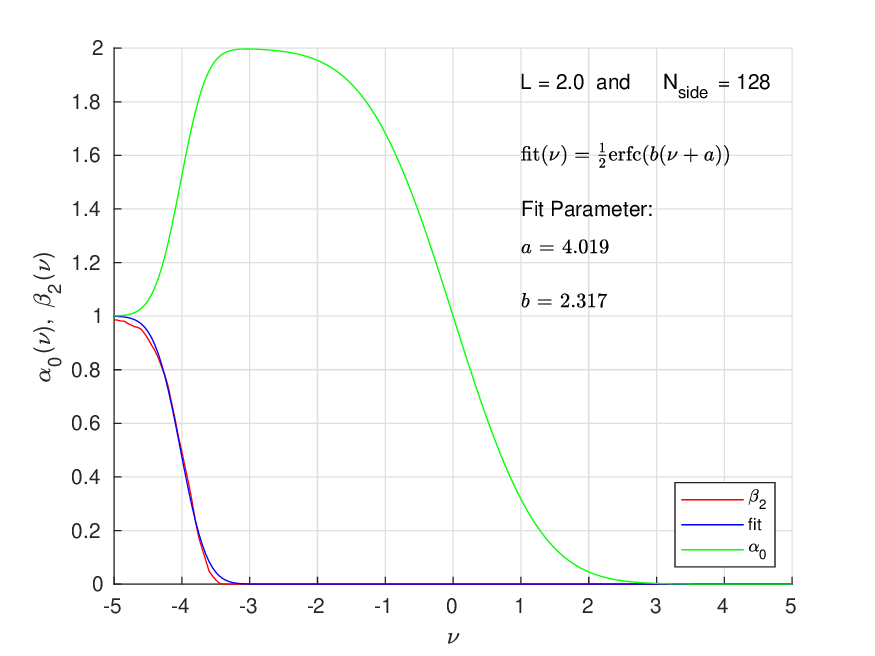}
\caption{\label{fig:beta2_alpha0}
  The mean value of $\beta_2(\nu)$ obtained from 1000 CMB simulations based
  on the torus topology is shown together with the fit (\ref{beta2_appr}).
  Also shown is $\alpha_0(\nu)$ obtained from (\ref{alpha0_appr}).
}
\end{figure}

Inserting (\ref{Minkowski_Gauss}) into (\ref{chi_MF}),
the exact Gaussion expectation for the EC reads
\begin{equation}
\label{chi_Gauss}
\chi(\nu) \; = \; \frac{\rho^2}{\sqrt{2\pi}} \, \nu \, e^{-\nu^2/2} \, + \,
\hbox{erfc}\left(\frac\nu{\sqrt{2}}\right)
\hspace{10pt} ,
\end{equation}
which compared with the approximation (\ref{chi_approx}) gives
\begin{equation}
\label{c1_Gauss}
c_1 \; = \; \frac{\rho^2}{\sqrt{2\pi}}
\end{equation}
and
\begin{equation}
\label{alpha0_Gauss}
\alpha_0(\nu) \; = \;
\hbox{erfc}\left(\frac\nu{\sqrt{2}}\right) \, - \, \beta_2(\nu)
\hspace{10pt} .
\end{equation}
In Fig.\,\ref{fig:beta2_alpha0} we show the mean value of $\beta_2(\nu)$
obtained from 1000 CMB simulations based on the torus topology and
compare it with the approximation
\begin{equation}
\label{beta2_appr}
\beta_2(\nu) \; = \; \frac 12 \hbox{ erfc}\Big(b(\nu+a)\Big)
\hspace{10pt} .
\end{equation}
Note that the fit parameters $a,b > 0$ have a simple interpretation since
$$
\beta_2(-a) \; = \; \frac 12
$$
and
$$
\beta_2'(-a) \; = \; -\, \frac{b}{\sqrt\pi}
\hspace{10pt} .
$$
Inserting (\ref{beta2_appr}) into (\ref{alpha0_Gauss})
leads to the analytic approximation
\begin{equation}
\label{alpha0_appr}
\alpha_0(\nu) \; = \; \hbox{erfc}\left(\frac\nu{\sqrt{2}}\right)
                     \, - \, \frac 12 \hbox{ erfc}\Big(b(\nu+a)\Big)
\hspace{10pt} ,
\end{equation}
also shown in Fig.\,\ref{fig:beta2_alpha0},
and thus to an explicit expression for the parity violation,
see eq.\,(\ref{parity_violation}).

\begin{figure}
\includegraphics[width=0.5\textwidth]{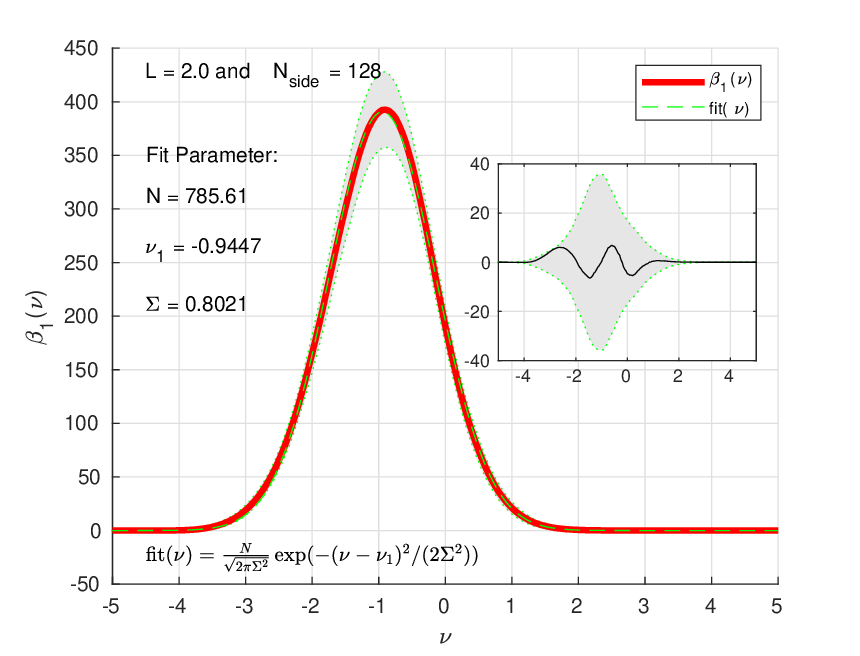}
\caption{\label{fig:beta1_fit}
  The mean value of $\beta_1(\nu)$ is shown together with a shifted Gaussian.
  The inset displays the difference between the Gaussian fit and $\beta_1(\nu)$
  in addition to the $1\sigma$ band confirming
  that the Gaussian fit decribes the data better than $1\sigma$.
}
\end{figure}

\begin{figure}
\includegraphics[width=0.5\textwidth]{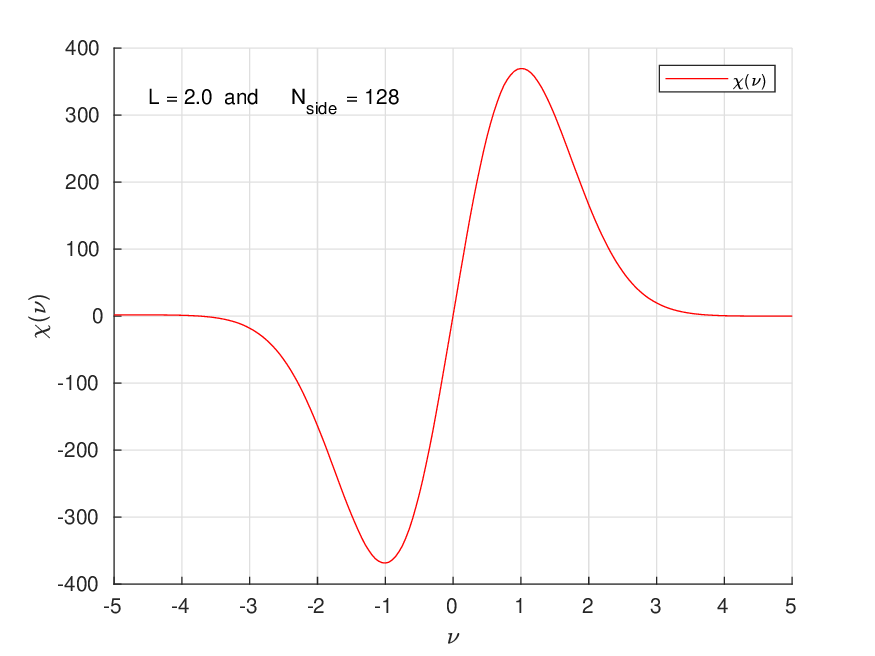}
\caption{\label{fig:euler}
The EC $\chi(\nu)$, eq.\,(\ref{Euler_beta}) is shown.
}
\end{figure}

To the best of our knowledge, Fig.\,\ref{fig:beta2_alpha0} presents for
the first time a computation of the BF $\beta_2(\nu)$ and of the function
$\alpha_0(\nu)$ quantifying the breaking of parity symmetry,
defined in eq.\,(\ref{parity_violation}).
In previous papers $\beta_2$ was considered to be zero and $\alpha_0(\nu)$
was not discussed at all.
From the definition (\ref{Def_Betti2}) it follows that $\beta_2(\nu)$ is
small since it is bounded by
\begin{equation}
\label{beta2_property}
0 \leq \beta_2(\nu) \leq 1
\hspace{5pt} \hbox{with} \hspace{5pt}
\beta_2(\nu) = 0
\hspace{5pt} \hbox{for} \hspace{5pt}
\nu > - \nu_2 \; \; (\nu_2 >0)
\hspace{5pt} .
\end{equation}
Also the parity violation $\alpha_0(\nu)$ is small
since it follows from (\ref{alpha0_Gauss}) and (\ref{beta2_property})
\begin{equation}
\label{alpha0_property}
\alpha_0(\nu) \; \leq \;
\hbox{erfc}\left(\frac\nu{\sqrt{2}}\right) \; \leq \; 2
\end{equation}
resp.
\begin{equation}
\label{alpha0_property_restricted}
\alpha_0(\nu) \; = \; \hbox{erfc}\left(\frac\nu{\sqrt{2}}\right)
\hspace{10pt} \hbox{for} \hspace{10pt}
\nu > - \nu_2
\hspace{10pt} .
\end{equation}
In contrast, the Figs.\ \ref{fig:beta_even}, \ref{fig:beta_odd},
\ref{fig:beta1_fit} and \ref{fig:euler} clearly show
that the $\beta_0(\nu)$, $\beta_1(\nu)$ and the EC $\chi(\nu)$
have very large maxima (minima) which finds a nice explanation
by the large value of the parameter $\rho$.
Indeed, we obtain from (\ref{coeff_c1}) and (\ref{c1_Gauss})
for the difference
\begin{equation}
\label{Def_delta1}
\delta_1(\nu) \; := \; \beta_1(-\nu) - \beta_1(\nu)
\end{equation}
at the maximum at $\nu=1$ (minimum at $\nu=-1$)
\begin{equation}
\label{delta1_extrema}
\delta_1(\pm 1) \; = \; \pm \frac{\rho^2}{\sqrt{2\pi\,\hbox{e}}}
\; = \; \pm 0.2420 \, \rho^2
\hspace{10pt} .
\end{equation}
In \cite{Aurich_Buchert_France_Steiner_2021} it was shown that the mean value
of $<\rho(L)>$ can be well approximated for tori of size $1.0\leq L \leq 3.0$
by the linearly decreasing function
\begin{equation}
\label{Behaviour_Rho}
<\rho(L)> \; \approx \; 46.122 \, - \, 3.290 \, L
\hspace{10pt} .
\end{equation}
From this we obtain for example for the torus with $L=2.0$ the
large value $<\rho(2)> \approx 39.542$ which leads to
$\delta_1(\pm 1) \approx \pm 378$ which explains the large amplitudes
displayed in Fig.\,\ref{fig:beta_odd}.
(A similar prediction follows for the BF $\beta_1(\nu)$,
see eqs.\,(\ref{beta1_Gaussian}), (\ref{beta1_0_rho}) and
Fig.\,\ref{fig:beta_even}.)
Thus, the BFs $\beta_0(\nu)$ and $\beta_1(\nu)$ can be used to detect
the size of the Universe if it is modelled as a cubic 3-torus.
See also Figs.\ \ref{fig:rho_L_minmax}, \ref{fig:/beta_L_sequence_min}
and \ref{fig:/beta_L_sequence_max}.

The figure \ref{fig:beta1_fit} demonstrates that a simple shifted Gaussian
\begin{eqnarray}
\label{beta1_Gaussian}
\beta_1(\nu) & = & \frac{N}{\sqrt{2\pi}\Sigma} \,
\exp\left(-\frac{(\nu-\nu_1)^2}{2\Sigma^2} \right)
\\ & = & \nonumber
\beta_1(0) \, \exp\left(-\frac{\nu(\nu-2\nu_1)}{2\Sigma^2} \right)
\end{eqnarray}
fits the mean value of $\beta_1(\nu)$ better than the $1\sigma$ width of the
distribution.
The normalization $N$, the width $\Sigma$ of the Gaussian and
the position $\nu_1$ of its maximum lead to the relations
\begin{eqnarray}
\label{beta1_Gaussian_b}
\beta_1(0) & = & \frac{N}{\sqrt{2\pi}\Sigma} \,
\exp\left(-\frac{\nu_1^2}{2\Sigma^2} \right)
\\ & = & \nonumber
\beta_1(\nu_1) \, \exp\Big(-\nu_1^2/(2\Sigma^2)\Big) \,
\; = \; \beta_1(2\nu_1)
\hspace{10pt} .
\end{eqnarray}
Furthermore, one derives from (\ref{beta1_Gaussian})
\begin{equation}
\label{beta1der_a}
\beta_1'(0) \; = \; \frac{\nu_1}{\Sigma^2} \, \beta_1(0)
\hspace{10pt} ,
\end{equation}
and from (\ref{coeff_c1}) and (\ref{c1_Gauss}) 
\begin{equation}
\label{beta1der_b}
\beta_1'(0) \; = \; - \,\frac{\rho^2}{2\sqrt{2\pi}}
\hspace{10pt} .
\end{equation}
Comparing the relations (\ref{beta1der_a}) and (\ref{beta1der_b})
yields the relation
\begin{equation}
\label{beta1_0_rho}
\beta_1(0) \; = \; \frac{\Sigma^2}{2\sqrt{2\pi}\,|\nu_1|} \; \rho^2
\hspace{10pt} ,
\end{equation}
which explains (see also (\ref{beta1_Gaussian_b}) and Fig.\,\ref{fig:euler})
the large amplitude of $\beta_1(\nu)$ due to the large value of $\rho^2$.

Finally, Fig.\,\ref{fig:euler} shows the EC for 1000 simulations
using the exact Gaussian expectation value (\ref{chi_Gauss}) of
the general relation (\ref{Euler_beta}).
Eq.\,(\ref{chi_Gauss}) predicts two large extrema
(for the $\rho$-values obtained from  (\ref{Behaviour_Rho})) at
\begin{equation}
\label{Def_nu_pm}
\nu_\pm \; := \; \pm \sqrt{1-\frac{2}{\rho^2}} \; \approx \; \pm 1
\end{equation}
with magnitude $\chi(\nu_+) \approx 379$ and $\chi(\nu_-) \approx -377$,
respectively;
for the torus with $L=2$ one has $\nu_\pm=\pm 0.999$.
It is interesting to note that even in the case where the primordial
initial conditions are exactly Gaussian,
there is a small ``parity breaking'' (not visible in Fig.\,\ref{fig:euler})
of the antisymmetry (negative parity) of $\chi(\nu)$ since it holds
\begin{equation}
\chi(-\nu) \; = \; -\chi(\nu) \, + \, 2
\hspace{10pt} .
\end{equation}

Another important topological quantity is the genus $g(\nu)$ related
to the EC by
\begin{equation}
\label{genus}
g(\nu) \; = \; 1 \, - \, \frac 12 \, \chi(\nu)
\hspace{10pt} .
\end{equation}
Inserting (\ref{chi_Gauss}) into (\ref{genus}),
one obtains for the Gaussian prediction of the genus the exact relation
\begin{equation}
\label{genus_Gauss}
g(\nu) \; = \; -\, \frac{\rho^2}{2\sqrt{2\pi}} \, \nu \, e^{-\nu^2/2} \, + \,
\frac 12 \,\hbox{erf}\left(\frac\nu{\sqrt{2}}\right) \, + \, \frac 12
\end{equation}
and
\begin{equation}
\label{genus_parity}
g(-\nu) \; = \; -\, g(\nu) \, + \, 1
\hspace{10pt} .
\end{equation}    
From (\ref{genus_Gauss}) one obtains the specical values
$g(\infty) = 1$ (torus), $g(0)=\frac 12$ and $g(-\infty) = 0$ (sphere).

\section{Betti Functionals for the cubic 3-torus topology}
\label{Sec:Betti_Functionals_from_Simulations}

In this section the properties of $\beta_0(\nu)$ and $\beta_1(\nu)$ are
discussed for the torus simulations with different torus side-lengths $L$.
Since the focus is put on the simulations, one does not has to bother about
masked sky regions,
which will be discussed in the next section.

\begin{figure}
\includegraphics[width=0.5\textwidth]{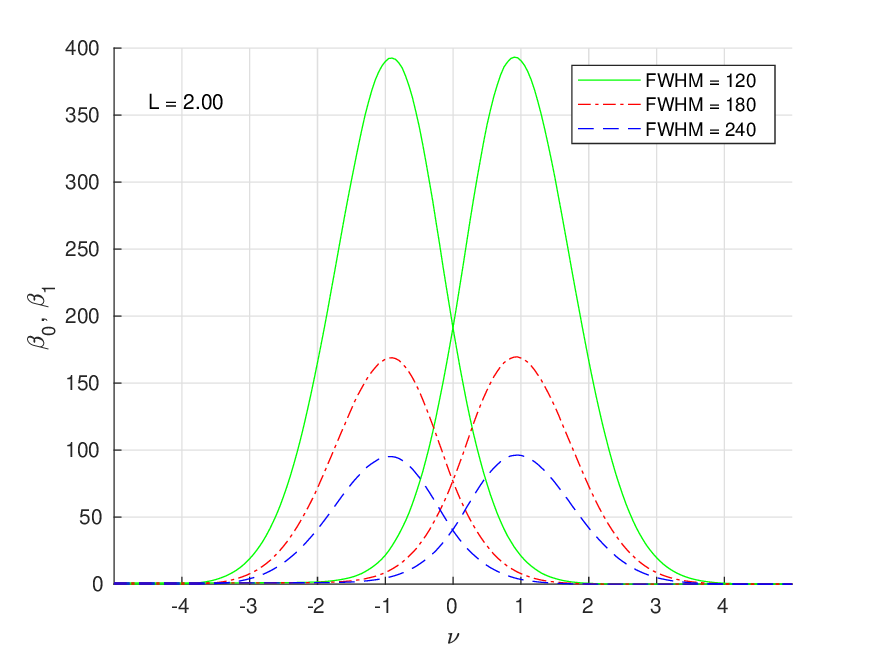}
\caption{\label{fig:beta_fwhm_sequence}
  The ensemble averages of $\beta_0(\nu)$ and $\beta_1(\nu)$ are shown
  for the torus side-length $L=2.0$ in dependence on the Gaussian smoothing
  of the sky maps.
  The smoothing of $2^\circ = 120\hbox{ arcmin}$, $3^\circ = 180\hbox{ arcmin}$,
  and $4^\circ = 240\hbox{ arcmin}$ are displayed.
}
\end{figure}

The figure \ref{fig:beta_fwhm_sequence} shows the influence of the
Gaussian smearing of the sky maps on $\beta_0(\nu)$ and $\beta_1(\nu)$.
With increasing smearing the excursions set has fewer structure elements
and thus, the amplitude of $\beta_0(\nu)$ and $\beta_1(\nu)$ decreases
with increasing smoothing.
This behaviour is nicely revealed in figure \ref{fig:beta_fwhm_sequence}
for the case $L=2.0$.
Since the 3-torus simulations of the sky maps are computed up to $\lmax =256$,
which roughly corresponds to a resolution of $180^\circ/\lmax \simeq 0.7^\circ$,
a Gaussian smoothing of at least $2^\circ$ is on the safe side.
As already stated, the analysis in this paper is based on a Gaussian smoothing
of $2^\circ$.

\begin{figure}
\includegraphics[width=0.5\textwidth]{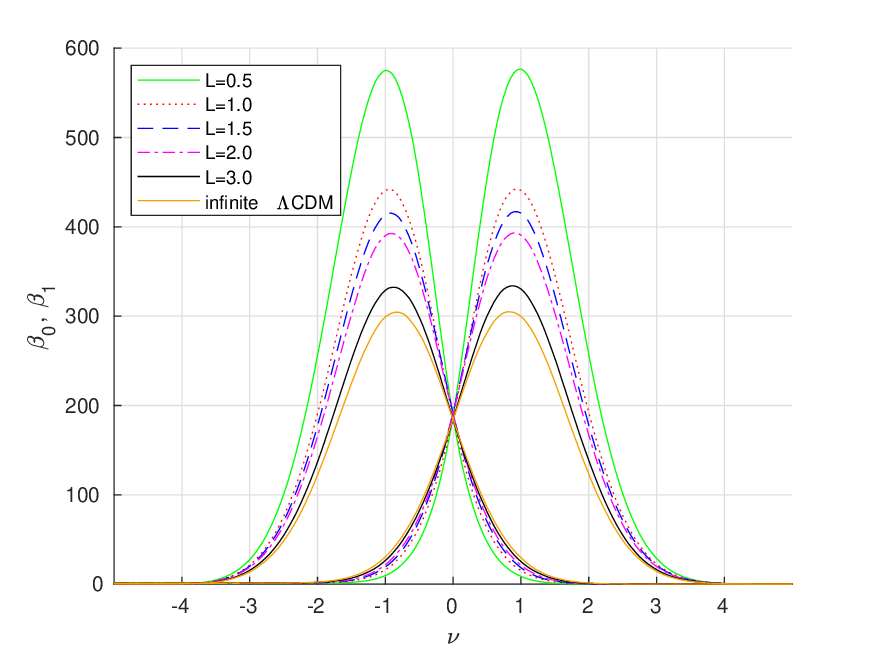}
\caption{\label{fig:beta_L_sequence}
  The ensemble averages of $\beta_0(\nu)$ and $\beta_1(\nu)$ are shown
  depending on the torus side-length for $L=0.5$ up to $L=3.0$.
  In addition, the ensemble averages of $\beta_0(\nu)$ and $\beta_1(\nu)$
  for the infinite $\Lambda$CDM model are shown,
  which extrapolates the torus side-length towards infinity.
  A decreasing number of structure elements with increasing side-length $L$
  is revealed.
  A Gaussian smoothing of $120\hbox{ arcmin}$ is used in all cases.
}
\end{figure}

We now compare for this fixed Gaussian smoothing the dependence of
the number of components and holes in dependence on the
side-length $L$ of the 3-torus in figure \ref{fig:beta_L_sequence}.
One observes a nice decreasing behaviour of $\beta_0(\nu)$ and $\beta_1(\nu)$
with increasing 3-torus size.
The figure \ref{fig:beta_L_sequence} also displays the corresponding result
for the infinite $\Lambda$CDM model,
which is computed using CAMB for the same set of cosmological parameters
as used for the 3-torus models.
This extrapolates the 3-torus size towards infinity.
So this monotone dependence on the size of the 3-torus might give a hint on
the size of our Universe by studying $\beta_0(\nu)$ and $\beta_1(\nu)$
in observed sky maps.

\begin{figure}
\includegraphics[width=0.5\textwidth]{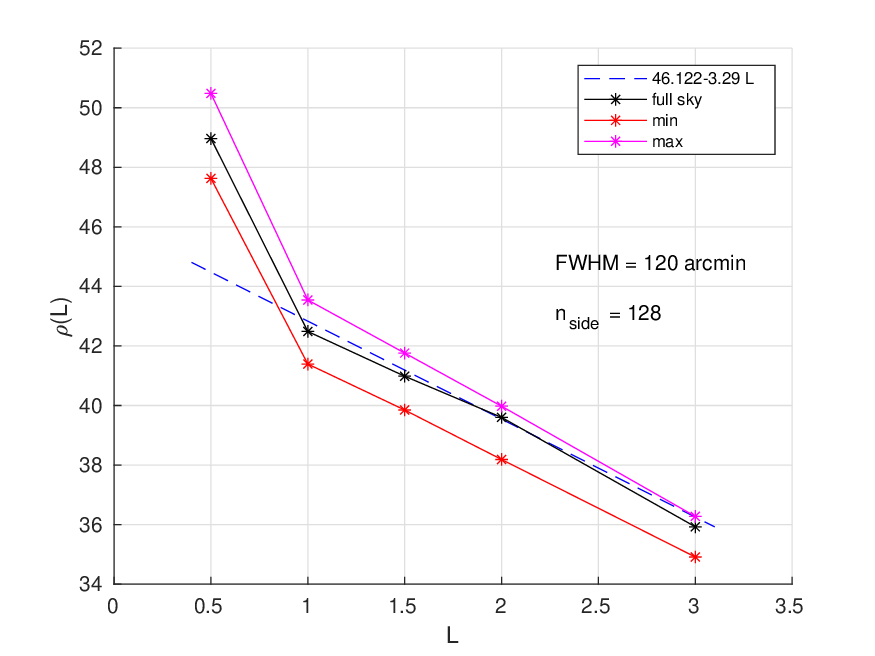}
\caption{\label{fig:rho_L_minmax}
  The dashed line shows the fit (\ref{Behaviour_Rho}) for $\rho$ determined in
  \cite{Aurich_Buchert_France_Steiner_2021},
  while the three full curves display the results obtained from $\beta_1(\nu)$
  using (\ref{rho_from_beta1}),
  where $\beta_1(\nu)$ of the full sky simulation and
  $\beta_1^{\hbox{\scriptsize min}}(\nu)/f_{\hbox{\scriptsize sky}}$ and
  $\beta_1^{\hbox{\scriptsize max}}(\nu)/f_{\hbox{\scriptsize sky}}$ are used.
  The latter ones are determined on the masked sky as explained in
  section \ref{section:masked_sky}.
}
\end{figure}

An interesting connection of this monotonic dependence on the size
of the topological cell exists with respect to
the normalized standard deviation of the CMB gradient field,
i.\,e.\ of $\rho$ already  introduced in
section \ref{sec:analytic_approximation}.
Solving relation (\ref{delta1_extrema}) with respect to $\rho$ yields
\begin{equation}
  \label{rho_from_beta1}
  \rho \; = \;
  \sqrt{\sqrt{2\pi\,\hbox{e}}\, \big(\beta_1(-1)-\beta_1(1)\big)}
  \hspace{10pt} .
\end{equation}
The interesting point is that this relation shows that a monotonic dependence
of $\beta_1$ on the torus size leads to an analogous behaviour of $\rho$.
While $\beta_1(\pm 1)$ is computed by counting the number of holes of
the excursion set,
$\rho$ is computed by differentiating the CMB temperature field.
In \cite{Aurich_Buchert_France_Steiner_2021} the behaviour of the mean value
of $<\rho(L)>$ is analyzed and a monotonic behaviour of $\rho$ is found,
which can be approximated for tori of size $1.0\leq L \leq 3.0$
by the already given linearly decreasing function (\ref{Behaviour_Rho}).
In figure \ref{fig:rho_L_minmax} the linear behaviour (\ref{Behaviour_Rho})
determined in \cite{Aurich_Buchert_France_Steiner_2021} is compared
with that derived from $\beta_1(\pm 1)$ by using (\ref{rho_from_beta1}).
A nice agreement between both methods for the computation of $\rho$
is observed.
The figure \ref{fig:rho_L_minmax} also shows the result by using
$\beta_1^{\hbox{\scriptsize min}}(\pm 1)/f_{\hbox{\scriptsize sky}}$ and
$\beta_1^{\hbox{\scriptsize max}}(\pm 1)/f_{\hbox{\scriptsize sky}}$,
which will be defined in the next section.

\section{Betti Functionals in the Presence of Masks}
\label{section:masked_sky}

\begin{figure}
\includegraphics[width=0.5\textwidth]{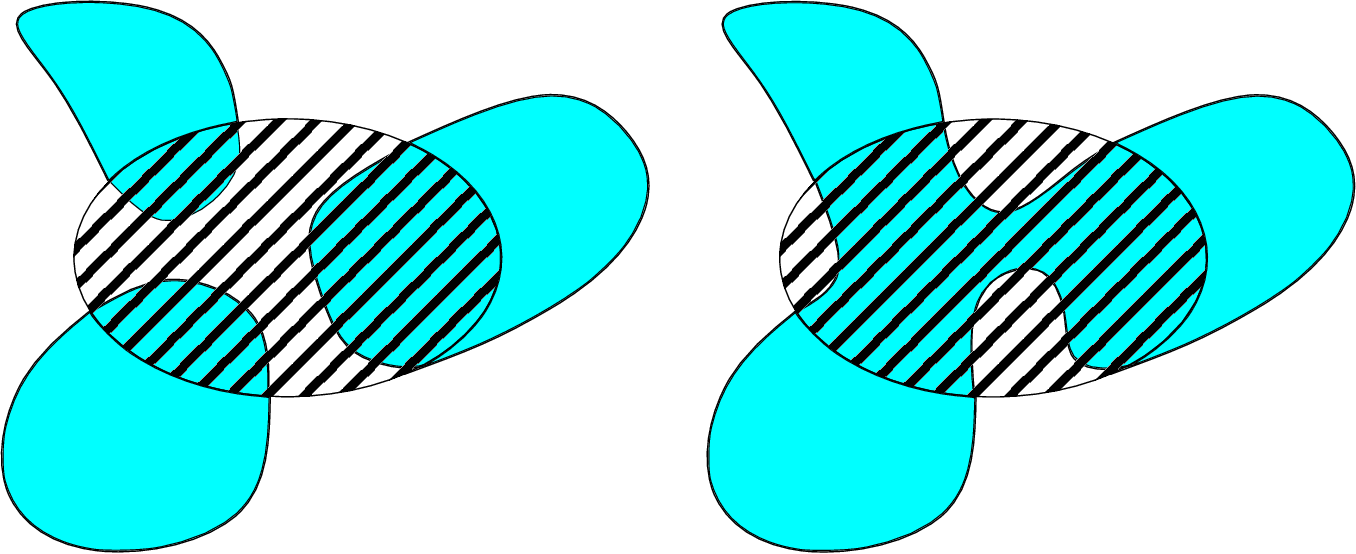}
\caption{\label{fig:min_max_with_mask}
  The ambiguity of counting the number of structure elements in the presence
  of a mask is illustrated.
  The mask is pictured as the dashed region,
  while the structure elements are shown as blue regions.
  If only the parts outside the masked region are known,
  one cannot decide whether they are connected within the mask or not.
  The two extreme cases are that they are all connected (right)
  or none of them (left).
}
\end{figure}

The analysis of the Betti functionals in CMB observations is impeded
by foregrounds that do not allow a measurement of the genuine CMB.
Furthermore, in the case of ground based observations one has to deal
with an incomplete sky coverage.
In order to estimate the Betti numbers in the case of a mask,
we propose the following procedure.
At first, for the computation of $\beta_0(\nu)$,
one counts the number of components in the unmasked sky.
Then there arises the possibility that some or all of those components
that are partially covered by a {\it common} connected masked region,
might be linked within this region
as illustrated in figure \ref{fig:min_max_with_mask}.
There three components are shown that might be connected within the
masked domain and
would be counted as one component in an ideally measured sky without a mask.
Of course, a further possibility is that only two of them are connected.
Therefore, a lower bound $\beta_0^{\hbox{\scriptsize min}}(\nu)$ is obtained
by counting all components that are touched by a common masked region
as a single component.
Conversely, an upper bound $\beta_0^{\hbox{\scriptsize max}}(\nu)$ is obtained
by treating all components as separated.
Note, that no attempt is made to estimate the number of components
within the masked domains,
since only components outside the masked regions are considered.
In order to estimate the number of components in the full sky,
$\beta_0^{\hbox{\scriptsize min}}(\nu)$ and $\beta_0^{\hbox{\scriptsize max}}(\nu)$
are divided by
\begin{equation}
\label{def:f_sky}
f_{\hbox{\scriptsize sky}} \; := \;
\frac{\hbox{Area of the unmasked sky}}{\hbox{Area of full sky}}
\hspace{10pt} .
\end{equation}
The same procedure applies analogously for $\beta_1$.

\begin{figure}
\includegraphics[width=0.5\textwidth]{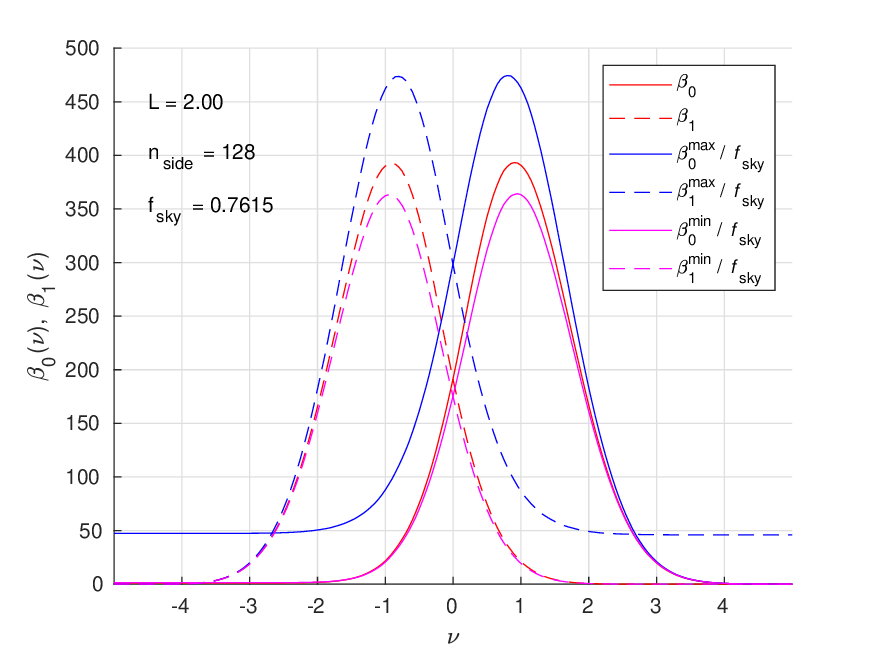}
\caption{\label{fig:beta_estimation_with_mask}
  The extrapolations
  $\beta_{0,1}^{\hbox{\scriptsize min}}(\nu)/f_{\hbox{\scriptsize sky}}$
  and $\beta_{0,1}^{\hbox{\scriptsize max}}(\nu)/f_{\hbox{\scriptsize sky}}$
  obtained from the masked CMB maps
  are shown in comparison with $\beta_{0,1}(\nu)$ derived from
  the unmasked CMB maps for the $L=2.0$ ensemble.
}
\end{figure}

In the case of simulations, one can test this procedure.
Figure \ref{fig:beta_estimation_with_mask} compares for the
case $L=2.0$ the upper and lower bounds with the
true full sky result.
The resolution parameters are again set to
$N_{\hbox{\scriptsize side}}=128$ and
$\hbox{FWHM}=120\hbox{ arcmin}$.

In this work, we use the Planck 2018
``{\it Component Separation Common mask in Intensity}''
\cite{Planck_2018_IV}
which can be obtained at
\url{http://pla.esac.esa.int/pla/#maps}
(file name: COM\_Mask\_CMB-common-Mask-Int\_2048\_R3.00.fits).
Since our analysis is based on the Healpix resolution
$N_{\hbox{\scriptsize side}}=128$,
we downgrade the above mask from
$N_{\hbox{\scriptsize side}}=2048$ to $N_{\hbox{\scriptsize side}}=128$.
The downgraded mask has no longer only the pixel values 0 and 1,
but also immediate values, and we use a mask threshold of 0.9
in the following analysis.
This leads to $f_{\hbox{\scriptsize sky}} = 0.7615$.

One observes from figure \ref{fig:beta_estimation_with_mask}
that the counted number of components lies nicely
between $\beta_0^{\hbox{\scriptsize min}}(\nu)/f_{\hbox{\scriptsize sky}}$ and
$\beta_0^{\hbox{\scriptsize max}}(\nu)/f_{\hbox{\scriptsize sky}}$, as it should be.
The same is seen for $\beta_1$ which refers to the holes.

Furthermore, for sufficiently low values of $\nu$,
$\beta_0^{\hbox{\scriptsize max}}(\nu)/f_{\hbox{\scriptsize sky}}$ saturates at
a non vanishing positive value.
This is due to the structure of the mask.
In the case without a mask, all CMB values are larger than $\nu$
for that sufficiently low $\nu$,
so that the full sphere $\Sspace^2$ is obtained as
the excursion set $\Excursion(\nu)$.
Applying the mask and assuming that the components are not connected
within the masked region, counts them as separate items
if they lie within a  ``hole'' of the mask.
Then each hole of the mask yields a separate component.
In contrast, $\beta_0^{\hbox{\scriptsize min}}(\nu)/f_{\hbox{\scriptsize sky}}$
assumes that all components are linked within the masked domains and
thus one counts only a single component.

\section{A comparison of the cubic 3-torus topology with the Planck CMB maps}
\label{section:Betti_and_Planck}

\begin{figure}
\includegraphics[width=0.5\textwidth]{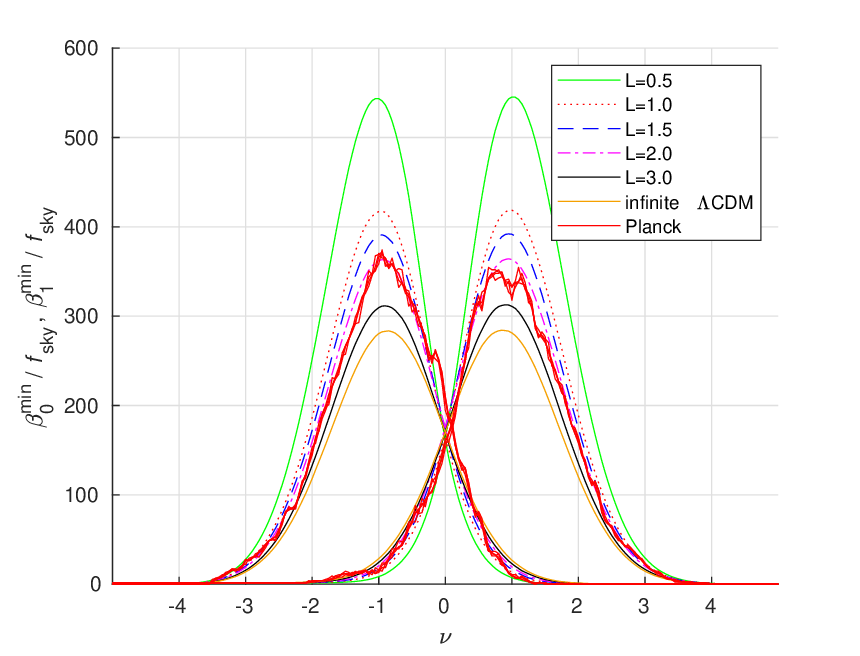}
\caption{\label{fig:/beta_L_sequence_min}
  $\beta_0^{\hbox{\scriptsize min}}(\nu)/f_{\hbox{\scriptsize sky}}$ and
  $\beta_1^{\hbox{\scriptsize min}}(\nu)/f_{\hbox{\scriptsize sky}}$ are displayed
  for the 3-torus simulations, the infinite $\Lambda$CDM model and
  the four Planck maps which are all subjected to the same mask.
}
\end{figure}

\begin{figure}
\includegraphics[width=0.5\textwidth]{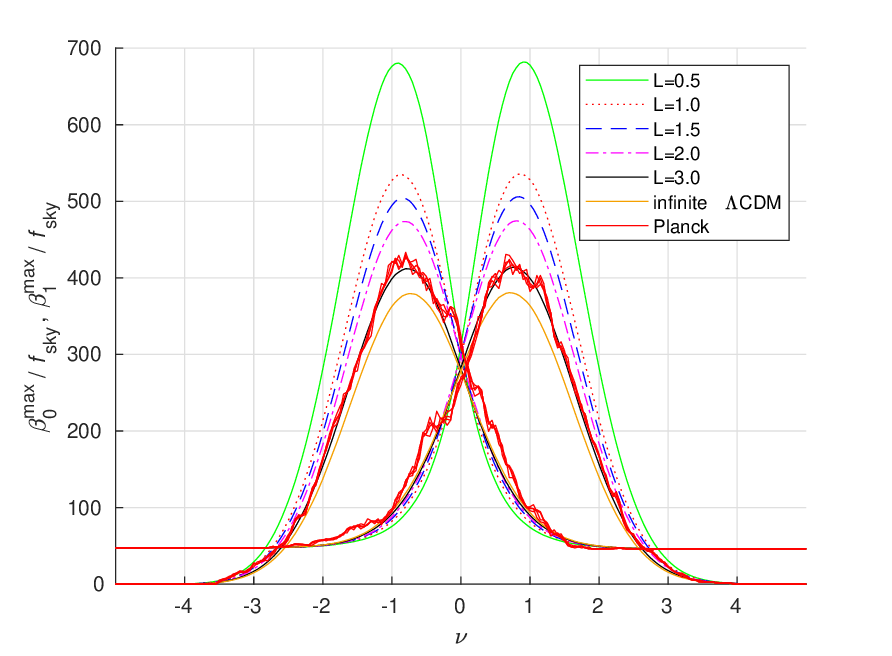}
\caption{\label{fig:/beta_L_sequence_max}
  The same as in figure \ref{fig:/beta_L_sequence_min}, but for
  $\beta_0^{\hbox{\scriptsize max}}(\nu)/f_{\hbox{\scriptsize sky}}$ and
  $\beta_1^{\hbox{\scriptsize max}}(\nu)/f_{\hbox{\scriptsize sky}}$.
}
\end{figure}

The Planck collaboration provides four CMB sky maps for
a cosmological analysis.
Here we use the Planck 2018 maps \cite{Planck_2018_I}
called {\it SMICA}, {\it Commander}, {\it NILC}, and {\it SEVEM},
which can be obtained at \url{http://pla.esac.esa.int/pla/#maps}.
The Healpix routine ``map2alm\_iterative'' is used to compute
the spherical expansion coefficients for the four Planck maps by taking
the mask {\it Component Separation Common mask in Intensity} into account.
Thereafter, the monopole and dipole are set to zero and
a Gaussian smoothing of $120\hbox{ arcmin}$ is applied.
These spherical expansion coefficients are used to generate four
Healpix maps in the resolution of $\Nside=128$,
which can then be compared with the corresponding 3-torus CMB maps.
The figures \ref{fig:/beta_L_sequence_min} and \ref{fig:/beta_L_sequence_max}
display the curves derived from these four Planck curves as solid and
in the common red colour, since they are nearly indistinguishable.

In figure \ref{fig:/beta_L_sequence_min}
$\beta_0^{\hbox{\scriptsize min}}(\nu)/f_{\hbox{\scriptsize sky}}$ and
$\beta_1^{\hbox{\scriptsize min}}(\nu)/f_{\hbox{\scriptsize sky}}$ are plotted
for the 3-torus simulations, the infinite $\Lambda$CDM model and
the four Planck maps all subjected to the same mask.
The curves present thus the lower estimate of the true $\beta$'s.
It is seen that the curves derived from the Planck maps
possess a significantly larger amplitude than that of
the infinite $\Lambda$CDM model.
Indeed, they lie between the cubic 3-torus simulations of the side-lengths
$L=2.0$ and $L=3.0$,
whereas $L=2.0$ provides the better match.
A similar behaviour is seen in figure \ref{fig:/beta_L_sequence_max},
where $\beta_0^{\hbox{\scriptsize max}}(\nu)/f_{\hbox{\scriptsize sky}}$ and
$\beta_1^{\hbox{\scriptsize max}}(\nu)/f_{\hbox{\scriptsize sky}}$ are plotted.
In this case, the Planck derived curves are closer to the $L=3.0$ case.
Thus, in both cases there seems to be an indication of
a finite size of our Universe
corresponding to a size between $L=2.0$ and $L=3.0$.

\section{Discussion and Summary}

In the quest for the global topology of the Universe,
there have been suggested several methods to unveil the spatial structure
on its very largest scales.
In this paper, we focus on the Betti functionals
applied to the excursion sets $\Excursion(\nu)$,
equation (\ref{excursion_set}), of the CMB,
which possess in this case a simple geometrical interpretation.
With the CMB observed on a sphere $\Sspace^2$,
the excursion set decomposes this sphere with respect to the normalized
temperature threshold $\nu$ into components and holes.
There are three Betti functionals in this case.
The first one $\beta_0$ counts the number of connected components,
the second one $\beta_1$ the number of topological holes,
and finally, $\beta_2$ the number of two-dimensional cavities.
The latter takes the value one, if the excursion set is the complete sphere,
that is if the threshold $\nu$ is smaller than the lowest normalized temperature
on the CMB sphere,
otherwise $\beta_2$ is zero.
Since the Betti functionals focus only on the number of structure elements,
they are even simpler than the Minkowski functionals.
This is because the Minkowski functionals require the computation of
the area, the circumference and a curvature measure of the boundary
of the components \cite{Buchert_France_Steiner_2017}.
As discussed in section \ref{sec:analytic_approximation},
the Minkowski functionals allow a derivation of a relation connecting the
average of the normalized standard deviation of the CMB gradient field
called $\rho$ with $\beta_1$, see equations (\ref{delta1_extrema})
and (\ref{rho_from_beta1}),
if the CMB is assumed to be a homogeneous, isotropic Gaussian random field.
Thus, the properties of $\rho$ and the Betti numbers are not independent.

The common lore is that a spatially finite universe is betrayed by the
large scale behaviour of the CMB,
for example the suppression of the quadrupole moment $C_2$ or the low
power in the 2-point angular correlation function $C(\vartheta)$
above sufficiently large angles on the sky, typically above $60^\circ$.
Often overlooked, a suppression at significantly smaller angles
of the angular correlation function $C(\vartheta)$ is additionally seen
such that the amplitude of $C(\vartheta)$ for the 3-torus models
is below that of the infinite $\Lambda$CDM model,
see \cite{Aurich_Janzer_Lustig_Steiner_2007}.
It should be emphasized that this small angle suppression is also visible
in $C(\vartheta)$ obtained from the observed sky.

The definition of $\rho$ as a differential measure reveals its obvious
local nature, so that a topological signature on small scales exists
also for this quantity,
since a dependence of $\rho$ on the volume of the cubic 3-torus
was demonstrated in \cite{Aurich_Buchert_France_Steiner_2021},
see also figure \ref{fig:rho_L_minmax}.
In section \ref{Sec:Betti_Functionals_from_Simulations}
it is shown that $\beta_0(\nu)$ and $\beta_1(\nu)$ display a hierarchical
dependence of their amplitudes with respect to the side-length $L$
of the cubic 3-torus, see figure \ref{fig:beta_L_sequence},
such that the amplitudes increase with decreasing volume $V=L^3$.
This behaviour is in nice agreement with
the normalized standard deviation $\rho$ of the CMB gradient field.
It reveals the local structure in the excursion set via $\beta_1(\nu)$
at $\nu=\pm 1$, see (\ref{rho_from_beta1}).
However, since $\beta_0$ and $\beta_1$ are, of course,
not restricted to the thresholds $\nu=\pm 1$,
they provide a more comprehensive tool than $\rho$.

The computation of $\beta_k(\nu)$ from observational sky maps is hindered
due to the presence of masks.
The number of connected components and holes is then ambiguous
since it is not discernible whether they are connected within
the not measured parts, i.\,e.\ within the mask.
In section \ref{section:masked_sky} a method is suggested which
gives for their number a lower and an upper bound within the observed sky.
Finally, section \ref{section:Betti_and_Planck} applies this method
to four sky maps released by the Planck collaboration in 2018,
called {\it SMICA}, {\it Commander}, {\it NILC}, and {\it SEVEM}.
The comparison with the cubic 3-torus simulations shows
that the curves derived from the four Planck maps lie between the 
3-torus models with side-length $L=2.0$ and $L=3.0$,
see figures \ref{fig:/beta_L_sequence_min} and \ref{fig:/beta_L_sequence_max}.
So this measure gives a further hint
that our Universe has a non-trivial topology.


\section*{Acknowledgements}

We would like to thank Thomas Buchert, Martin France and Pratyush Pranav
for discussions.
The software packages HEALPix (\url{http://healpix.jpl.nasa.gov},
\cite{Gorski_Hivon_Banday_Wandelt_Hansen_Reinecke_Bartelmann_2005})
and CAMB written by A.~Lewis and A.~Challinor (\url{http://camb.info})
as well as the Planck data from 
\url{http://pla.esac.esa.int/pla/#maps} were used in this work.


\section*{References}

\bibliography{../bib_astro.bib}
\bibliographystyle{iopart-num-doi}

\end{document}